\documentclass[10pt]{iopart}

\usepackage{float}
\usepackage{epsfig}
\usepackage{cite}
\usepackage{caption}
\usepackage{iopams} 
\expandafter\let\csname equation*\endcsname\relax

\expandafter\let\csname endequation*\endcsname\relax
\usepackage{amsmath}
\begin{document}

\title{Dissociative electron attachment to sulfur dioxide : A theoretical approach}
\author{Irina Jana$^1$, Sumit Naskar$^2$, Mousumi Das$^{2,*}$ and Dhananjay Nandi$^{1,**}$}

\address{$^1$Department of Physical Sciences, Indian Institute of Science Education and Research Kolkata, Mohanpur 741246, India.}
\address{$^2$Department of Chemical Sciences, Indian Institute of Science Education and Research Kolkata, Mohanpur 741246, India.}
\ead{$^*$mousumi@iiserkol.ac.in; $^{**}$dhananjay@iiserkol.ac.in}
\vspace{10pt}

\begin{abstract}
	In this present investigation, density functional theory (DFT) and natural bond orbital (NBO) calculations have been performed to understand experimental observations of dissociative electron attachment (DEA) to SO$_2$. The molecular structure, fundamental vibrational frequencies with their corresponding intensities and molecular electrostatic potential (MEP) map, signifying the electron density contours, of SO$_2$ and SO$_2^-$ are interpreted from respective ground state optimized electronic structures calculated using DFT. The MEPs are then quantified and the second order perturbation energies for different oxygen lone pair (n) to $\sigma^*$ and $\pi^*$ interactions of S-O bond orbitals have been calculated by carrying out NBO analysis and the results are investigated. The change in the electronic structure of the molecule after the attachment of a low-energy ($\leq$ 15 eV) electron, thus forming a transient negative ion (TNI), can be interpreted from the $n\rightarrow\sigma^*$ and $n\rightarrow\pi^*$ interactions. The results of the calculations are used to interpret the dissociative electron attachment process. The dissociation of the anion SO$_2^-$ into negative and neutral fragments has been explained by interpreting the infrared (IR) spectrum and the different vibration modes. It could be observed that the dissociation of the anion SO$_2^{-}$ into S$^-$ occurs as a result of simultaneous symmetric stretching and bending modes of the molecular anion. While the formation of O$^-$ and SO$^-$ occurs as a result of anti-symmetric stretching of the molecular anion. The calculated symmetries of the TNI state contributing to the first resonant peak at around 5.2 eV and second resonant peak at around 7.5 eV could be observed from time-dependent density functional theory (TD-DFT) calculations to be an A$_1$ and a combination of A$_1$+B$_2$ states for the two resonant peaks, respectively. These findings strongly support our recent experimental observations for DEA to SO$_2$ using the sophisticated velocity map imaging (VMI) technique [Jana and Nandi, \textit{Phys. Rev. A, 2018, \textbf{97}, 042706}].
%
%
\end{abstract}

%
%
%

%
\ioptwocol

\section{Introduction}
Sulfur dioxide is a bent molecule with O-S-O bond angle of 119.3$^0$ and S-O bond length of 143.1pm having a C$_{2v}$ symmetry with ground state configuration (7a$_1$)$^2$,(1a$_2$)$^2$,(4b$_2$)$^2$,(8a$_1$)$^2$ \cite{EK1}. The presence of sulfur dioxide in starting from acid rain to natural volcanic sources, makes it one of the most important atmospheric molecules \cite{mcconkey}. Recording the vibrational bands and molecular constants of SO$_2$ from the infrared (IR) spectrum of the molecule using an infrared prism-grating spectrometer have been reported since the early 1950's \cite{shelton}. Robert \textit{et al.} also reported IR the  spectrum of anhydrous liquid sulfur dioxide and anhydrous liquid hydrogen fluoride using a Perkin-Elmer, Model 21, double beam recording infrared spectrometer \cite{robert}. Recently, in the year 2014, Zhu \textit{et al.} reported density functional theory (DFT) and natural bond orbital (NBO) calculations to study the electronic structures and bonding interactions between sulfur dioxide molecule and ruthenium (II) atom in two ruthenium-SO$_2$ adducts \cite{zhu2014bonding}. Using the molecular structure, vibrational spectra and molecular electrostatic  potential to study large polyatomic molecules like metolazone using the Gaussian 03 program package can be observed in the literature \cite{boopathi}. \textit{Ab initio} calculations to investigate the enhancement of halogen bonds by $\sigma$-hole and $\pi$-hole interactions between two molecules containing halogen atom in one, and a negative site in another has been reported by Esrafili and Vakili where, molecular electrostatic potential (MEP) of isolated SO$_2$ has been computed \cite{Vakili}. \\

Study of low-energy ($\leq$ 15 eV) electron attachment to SO$_2$ has been a topic of interest since the early 1970's. There have been many experimental studies reporting dissociative electron attachment (DEA) to SO$_2$ \cite{EK1,EK2,OJO,Rallis,Spyrou,Cadez_1983,jana}. Experiments on DEA to SO$_2$ is known to produce two prominent resonant peaks at around 5.2 and 7.5 eV as explained by the following reaction:
\begin{equation}
SO_2 + e^-\longrightarrow SO_2^{-*}\longrightarrow 
\begin{cases}
S^-+O_2\\
O^-+SO\\  
SO^-+O
\end{cases}
\label{eqn1}
\end{equation}
where a low-energy electron gets attached to the SO$_2$ molecule via a resonant capture forming SO$_2^{-*}$. This complex SO$_2^{-*}$, called the transient negative ion (TNI), then dissociates giving neutral and anionic fragments.\\

Experiments reporting electron attachment cross-section of the negative ion fragments produced from DEA to SO$_2$ and also the corresponding kinetic energies of the fragments using different processes are abundant in the literature \cite{Rallis,OJO,Cadez_1983,EK2,Spyrou}. Although experiments reporting angular distribution of the fragments are scarce \cite{gope}. But theories reporting DEA to SO$_2$ are very less in the literature \cite{gupta-baluja,EK1}. In the year 1996, Krishnakumar \textit{et al.} using \textit{ab initio} molecular orbital calculations and selection rules for dissociative electron attachment reported the first resonance between 4-5 eV to be due to $^2A_1$ and the second peak at around 7 eV to be due to $^2B_2$ negative ion resonant states \cite{EK1}. Recently, Jana and Nandi reported DEA to SO$_2$ using the sophisticated velocity map imaging (VMI) and identified the first resonance around 5.2 eV due to an $A_1$ and the second resonance around 7.5 eV due to a combination of $A_1$+$B_2$ negative ion resonant states \cite{jana}.\\

We report a theoretical and computational approach to find out the optimized ground state molecular structure of neutral SO$_2$ molecule and its anionic part SO$_2^-$ formed after the attachment of a low energy electron to the molecule using Gaussian 09 program package \cite{gauss}. The density functional theory (DFT) calculation has been performed using the Becke 3LYP exchange-correlational functional containing the Slater exchange functional, Hartree-Fock and Becke's 1988 gradient correction, along with the Lee-Yang-Parr functional with a aug-cc-pVQZ basis set to obtain a stable structure \cite{sumit,becke,Lee,becke2,lyp,kendall1992electron-aug-cc-pvnz}. The results of these calculations are then used to find out change in the ground state electronic structure of the molecule after the electron attachment. The electron density distribution is then visualized for the neutral SO$_2$ molecule and the SO$_2^-$ TNI with the help of the molecular electrostatic potential (MEP) map. The electron density distribution has been quantified with the help of natural bond orbital (NBO) analysis, followed by bond order calculation and Mulliken charge analysis. A vibration spectral analysis is then carried out using the computed infrared (IR) spectrum, to investigate different vibration modes of the TNI and hence identify the formation of different negative and neutral fragments given by Eqn. \ref{eqn1}. The potential curves for the first two dissociation pathways of Eqn. \ref{eqn1} have also been plotted for the neutral parent molecule and the optimized ground state configuration of the TNI, by varying the O-O and S-O bond distances, respectively. Finally, a time-dependent density functional theory (TD-DFT) calculation has been performed to calculate the vertical excited state energies of SO$_2^-$ TNI and thus identify the negative ion resonant states of the TNI involved in DEA to SO$_2$ at the first and second resonant peaks.

\section{Modeling of the molecule and density functional theory}

Density functional theory (DFT) calculations to small polyatomic molecules is abundant in the literature. In the present work, \textit{ab initio} electronic structure calculation using DFT has been performed to neutral SO$_2$ and its anionic part SO$_2^{-}$ using Gaussian 09 program package \cite{gauss}. Quantum chemical calculations using DFT is an efficient tool in illuminating not only the electronic structure of the SO$_2$ molecule, but also how the neutral molecule reacts to the incoming low-energy electron thus forming the TNI state. It is well known that the electronic structure of anions can be obtained using DFT \cite{Rosch}. First, the ground state geometries of SO$_2$ and SO$_2^{-}$ molecules were optimized using DFT with the Becke 3LYP exchange-correlational functional with a aug-cc-pVQZ basis set to obtain a stable structure. The ground state optimized energy for SO$_2$ comes out to be -548.733 Hartree while the optimized energy for ground state of SO$_2^{-}$ comes out to be -548.785 Hartree. Thus the optimized ground state energy of SO$_2^{-}$ is 1.42 eV lower than the ground state optimized energy of the neutral molecule. This indicates that the adiabatic electron affinity of SO$_2$ is positive and given by the difference value of 1.42 eV. This matches excellently with Grabowski \textit{et al.} who measured the electron affinity of SO$_2$ to be 1.1$\pm$0.1 eV using the flowing afterglow technique \cite{EA_SO2}. \\

\begin{figure}[h]
	\centering	
	\includegraphics[width=0.7\linewidth]{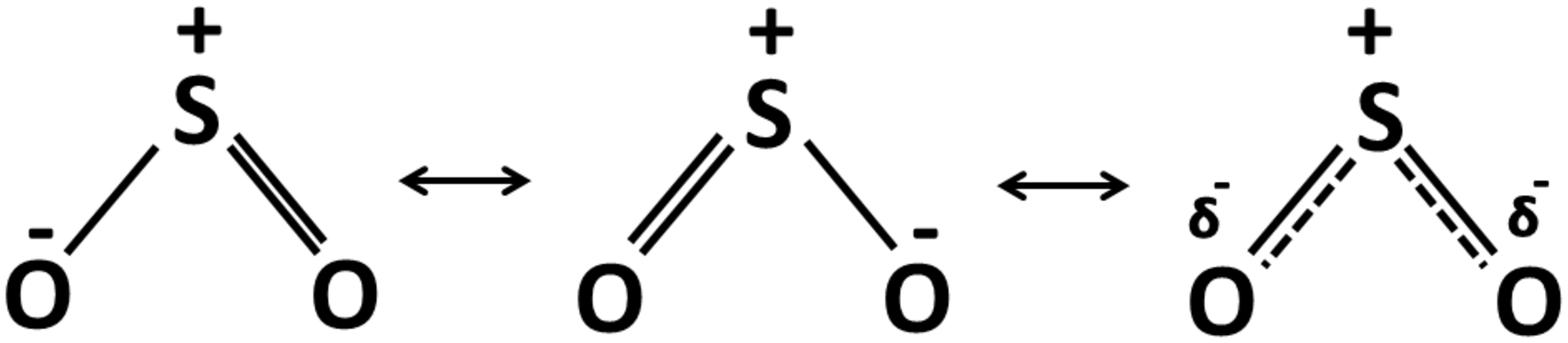}
	\caption{\label{so2_lewis} Lewis structure of SO$_2$ molecule. }
\end{figure}

Considering the Lewis structure of SO$_2$ molecule (Fig. \ref{so2_lewis}) it can be observed that the bent structure of the molecule is such that the S-atom has a less electro-negativity while the bottom O-atoms have more of it. This makes the SO$_2$ molecule polar. This charge difference and the distance between the charge centers produces a large permanent dipole moment. The dipole moment of SO$_2$ is computed to be 1.7473 D while the theoretical dipole moment reported by McConkey \textit{et al.} is 1.63305 D \cite{mcconkey}. Thus the dipole moment of the neutral molecule matches satisfactorily with the earlier reported value. As a low-energy electron gets attached to the SO$_2$ molecule forming a TNI, the charge distribution of the molecule as a whole changes and so does the dipole moment. The dipole moment of the optimized SO$_2^{-}$ ground state is observed to be 1.4481 D. This 0.30 D decrease can be attributed to the attachment of the negative charge.\\

\begin{figure}[h]
	\centering	
	\includegraphics[width=0.7\linewidth]{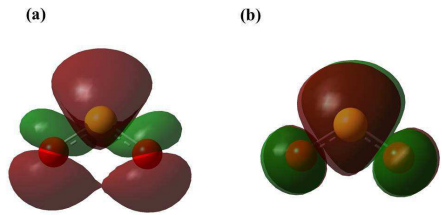}
	\caption{\label{homo_lumo_ex_pVQZ} (a) 3D visualization of the HOMO and (b) 3D visualization of the LUMO orbitals of SO$_2^-$ molecule. The yellow sphere represents the S-atom and the red spheres represent the O-atoms. }
\end{figure}

The present calculation recognizes the SO$_2$ ground state to be an $^1$A$_1$ state while the SO$_2^{-}$ as a $^2$B$_1$ state. This identification is in well-agreement with Gupta and Baluja \cite{gupta-baluja}. The highest occupied molecular orbital (HOMO) and lowest unoccupied molecular orbital (LUMO) for these configurations are 8a$_1$ and 9b$_1$ respectively, for both SO$_2$ and SO$_2^{-}$ molecules, as shown in Fig. \ref{homo_lumo_ex_pVQZ}. The gap between the HOMO and LUMO for SO$_2$ ground state optimized geometry is 5.12 eV while, for SO$_2^{-}$ ground state optimized geometry it is 2.78 eV. The HOMO of SO$_2^{-}$ rises by an amount of 9.97 eV and the LUMO of SO$_2^{-}$ rises by an amount of 7.63 eV as compared to the respective HOMO and LUMO values of SO$_2$, making the TNI SO$_2^{-}$ more stable. \\

\begin{figure}[h]
	\centering	
	\includegraphics[width=0.65\linewidth]{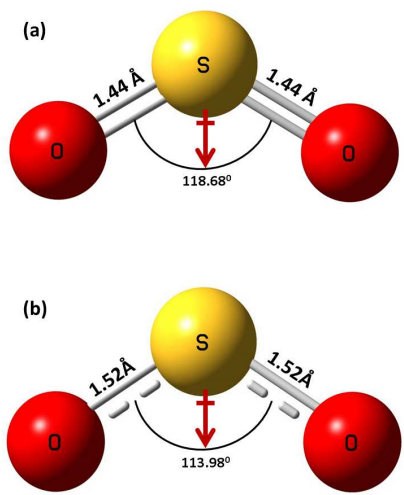}
	\caption{\label{SO2} The calculated optimized ground state geometries of: (a) SO$_2$ and (b) SO$_2^-$ ion both showing the C$_{2v}$ symmetry. The red arrow denotes the direction of the dipole moment for both the cases.}
\end{figure}

The calculated geometries of the SO$_2$ molecule and SO$_2^-$ negative ion are shown in Fig. \ref{SO2}(a) and Fig. \ref{SO2}(b), respectively. The neutral SO$_2$ molecule has C$_{2v}$ symmetry with a S-O bond length of 1.44 $\mathring{A}$ and 118.68$^0$ O-S-O bond angle in the optimized state (Fig. \ref{SO2}(a)). After the attachment of the electron, the anion retains the C$_{2v}$ symmetry but the S-O bond length increases by 0.08 $\mathring{A}$ while the O-S-O bond angle decreases by 4.7$^0$ (Fig. \ref{SO2}(b)) \cite{jana}. While the direction of the dipole moment remains same in both the cases and is represented by the red arrow in Fig. \ref{SO2}(a) and Fig. \ref{SO2}(b). \\

\section{Molecular electrostatic potential}
The change in the dipole moment is associated with a change in the molecular charge distribution which can be visualized with molecular electrostatic potential (MEP) maps. When a charged particle comes in the vicinity of a molecule, the charge cloud generated through the molecule's electrons and nuclei may serve as a guide to assess the molecule's reactivity towards positively or negatively charged particle. Thus, the electrostatic potential maps can be used as an effective tool to accurately analyze the charge distribution of a molecule and correlate molecular properties like dipole moments, nucleophilic and electrophilic site and reactivity of the molecule towards charged reactants \cite{murray_book}.\\

\begin{figure}[h]
	\centering	
	\includegraphics[width=0.65\linewidth]{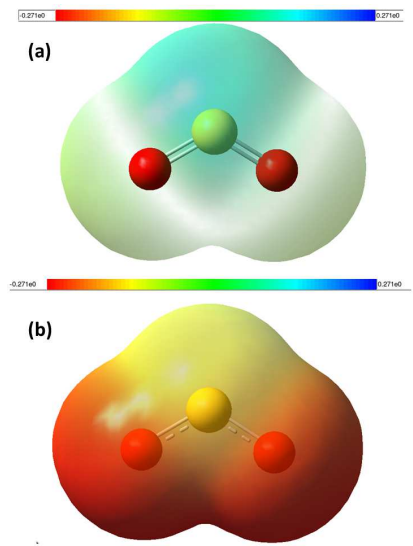}
	\caption{\label{MEP} Electrostatic potential maps at 0.001 electron/Bohr$^3$ isodensity surface of: (a) SO$_2$ molecule, (b)SO$_2^{-}$ ion. The red in the color spectrum denotes lowest electrostatic potential value while the blue denotes the highest. The yellow sphere represents the S-atom and the red spheres represent the O-atoms. }
\end{figure}

The optimized structural parameters are used to plot the MEP for SO$_2$ and SO$_2^{-}$ ground states as shown in Fig. \ref{MEP}(a) and Fig. \ref{MEP}(b), respectively. The red in the colour spectrum denotes lowest electrostatic potential value while the blue denotes the highest. The colour spectrum ranges from -0.271 a.u. (deepest red) to +0.271 a.u. (deepest blue) in both the figures (Fig. \ref{MEP}(a) and Fig. \ref{MEP}(b)). It can be observed from Fig. \ref{MEP}(a) that the region of lowest electrostatic potential energy, characterized by abundance of electrons, occurs near the bottom O-atoms which is in accordance with the Lewis structure (Fig. \ref{so2_lewis}) and also in well agreement with the MEP reported by Esrafili and Vakili \cite{Vakili}. From Fig. \ref{MEP}(b) it can be observed that after the attachment of the low-energy electron with the SO$_2$ molecule, the MEP map of SO$_2^{-}$ shifts towards higher electron density. But the region of lowest electrostatic potential still remains near the O-atoms. This points to the fact that although the electron gets attracted towards the electro-positive S-atom, it quickly gets transfered to the O-atoms. The TNI retains the C$_{2v}$ symmetry like its neutral ground state and the dipole moment vector can be identified to be pointing downwards in both the cases (Fig. \ref{SO2}(a) and Fig. \ref{SO2}(b)).\\

\begin{figure}[h]
	\centering	
	\includegraphics[width=0.65\linewidth]{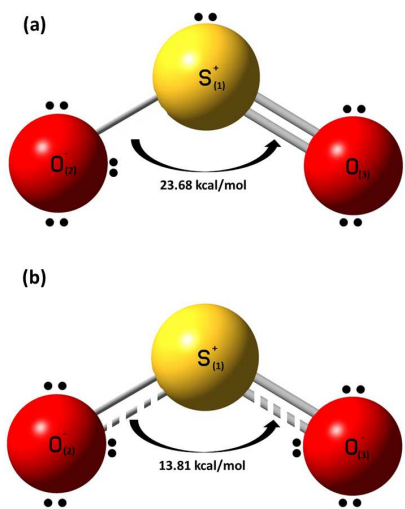}
	\caption{\label{nbo2} Schematic showing lone-pairs; $\sigma^*$ and $\pi^*$ orbitals with corresponding labels from NBO analysis for (a) neutral SO$_2$ molecule; (b) SO$_2^-$ ion. The small black dots denote the lone-pair electrons.}
\end{figure}

The bonding features between the incoming electron and the neutral SO$_2$ molecule can be investigated and the MEP can be quantified with the help of natural bond orbitals (NBO) analysis. The NBO 3.0 analysis, inbuilt within Gaussian 09 program package, has been carried out to observe the stabilization energies for donor-to-acceptor interactions. The NBO calculations for the SO$_2$ ground state show that the top S-atom (labeled S$_{(1)}$ in Fig. \ref{nbo2}(a)) contains a single lone pair of electrons with an occupancy of 1.993. While the two bottom O-atoms (labeled O$_{(2)}$ and O$_{(3)}$ in Fig. \ref{nbo2}(a), respectively) share three and two lone electron pairs each thus having a more electro-negativity as compared to the S-atom. The O-atom having three lone electron pairs (O$_{(2)}$) is observed with occupancies of 1.994, 1.843 and 1.545, respectively.  Whereas, the other O-atom with two lone electron pairs (O$_{(3)}$) has occupancies of 1.994 and 1.843 for the first and second lone pairs, respectively. Second order perturbation energy calculated for the lone pair of O$_{(2)}$ and $\pi^*$ bond orbital of S$_{(1)}$-O$_{(3)}$ ($n\rightarrow\pi^*$) interaction is 23.68 kcal/mol (Fig. \ref{nbo2}(a)). Since the $\sigma$ electrons form the molecular backbone, these electrons are tightly bound. The second order perturbation correction suggests that the $n\rightarrow\sigma^*$ interaction energy between the lone pair of O$_{(2)}$ and $\sigma^*$ bond orbital of S$_{(1)}$-O$_{(3)}$ is 76.84 kcal/mol which is more  than the $n\rightarrow\pi^*$ interaction. Thus the $n\rightarrow\pi^*$ interaction is energetically favorable. \\

As a low-energy electron gets attached to the neutral molecule forming the TNI SO$_2^{-}$, the charge distribution of the molecule completely changes as predicted by the NBO calculations. NBO calculations for the SO$_2^-$ ground state show that the top S$_{(1)}$-atom contains a single lone pair of electrons with an occupancy of 1.998. While the two bottom O$_{(2)}$ and O$_{(3)}$ atoms share three lone electron pairs each thus having a more electro-negativity as compared to the S$_{(1)}$-atom (Fig. \ref{nbo2}(b)). The O$_{(2)}$ and O$_{(3)}$ atoms each have occupancies of 1.995, 1.983 and 1.851 for the three lone pairs, respectively. Second order perturbation energy calculated for the $n\rightarrow\sigma^*$ interaction of second lone pair of O$_{(2)}$ and $\sigma^*$ bond orbital of S$_{(1)}$-O$_{(3)}$ is 13.81 kcal/mol because there is no $\pi^*$ orbital for SO$_2^{-}$ ground state (Fig. \ref{nbo2}(b)). The second order perturbation energy difference of 9.87 kcal/mol (23.68 - 13.81 kcal/mol) between the $n\rightarrow\pi^*$ for SO$_2$ and $n\rightarrow\sigma^*$ for SO$_2^{-}$ interaction suggests that the stabilization energy for SO$_2^{-}$ is much less than that for neutral SO$_2$. Thus, due to donor-acceptor interaction, SO$_2^{-}$ ground state is more stable than the SO$_2$ ground state. This is also evident from the HOMO-LUMO gap (5.12 eV for neutral SO$_2$ and 2.78 eV for SO$_2^{-}$). \\

\begin{figure}[h]
	\centering	
	\includegraphics[width=0.65\linewidth]{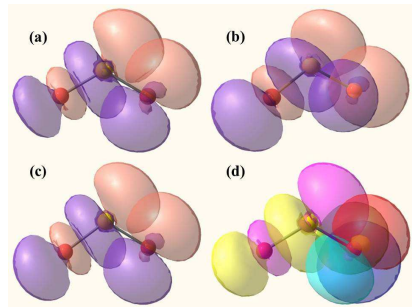}
	\caption{\label{nbo} The NBO density plot for SO$_2$ and SO$_2^{-}$ molecule. (a) Orbital density plot for SO$_2$ molecule representing the $n\rightarrow\pi^*$ interaction. (b) Orbital density plot for SO$_2^{-}$ molecule representing the $n\rightarrow\sigma^*$ interaction ($\alpha$ orbital). (c) Orbital density plot for SO$_2^{-}$ molecule representing the $n\rightarrow\sigma^*$ interaction ($\beta$ orbital). (d) Orbital density plot for SO$_2^{-}$ molecule representing the $n\rightarrow\sigma^*$ interaction ($\alpha$ and $\beta$ orbitals).}
\end{figure}

The NBO density plot for SO$_2$ and SO$_2^{-}$ has been shown in Fig. \ref{nbo}. Initially, for the ground state of SO$_2$ there is significant overlap between the lone pair of O$_{(2)}$ and $\pi^*$ bond orbital of S$_{(1)}$-O$_{(3)}$ (Fig. \ref{nbo}(a)). Since the single electron gets attached to the neutral SO$_2$ in up-spin state ($\alpha$ state), the NBO density plot representing the $\beta$ orbital of $n\rightarrow\sigma^*$ interaction remains unchanged (Fig. \ref{nbo}(c)). The presence of the $\alpha$ spin results in an extra overlap in the same LUMO orbital represented by the $n\rightarrow\sigma^*$ interaction solely due the extra electron. This can be observed from the NBO density plot for SO$_2^{-}$ alpha orbital (Fig. \ref{nbo}(b)). The resultant NBO density plot for SO$_2^{-}$ gives the orbital plot for a total sum of $\alpha$ and $\beta$ orbitals (Fig. \ref{nbo}(d)). Comparing the resultant NBO density plot for SO$_2^{-}$ to that for SO$_2$ neutral state, it can be seen that there is a much higher overlap in the anionic state than the neutral state. This higher overlap again implies the lower interaction energy as predicted by the NBO calculations.\\
It can be noted from Fig. \ref{SO2} that the optimized ground state configuration for the neutral SO$_2$ (Fig. \ref{SO2}(a)) and TNI SO$_2^-$ (Fig. \ref{SO2}(b)) are not identical. This change in the bond S-O length and O-S-O bond angle can be determined from the change in bond order using the following formula:\\

	Bond Order=$\dfrac{B_e + A_e}{2}$	\\
	where, $B_e$ and $A_e$ denotes number of bonding electrons and number of anti-bonding electrons, respectively.\\

 The bond order calculated using the bonding and anti-bonding electrons from the NBO analysis (relevant to Fig. \ref{nbo2}) are given in Table \ref{bond_order}. As can be noted from Fig. \ref{nbo2}(a), the resonating single S$_{(1)}$-O$_{(2)}$ and double S$_{(1)}$-O$_{(3)}$ bonds correspond to bond orders of 0.9 and 1.9, respectively. This means, for the optimized ground state configuration where the lone pairs of electrons redistribute themselves forming two double bonds (as in Fig. \ref{SO2}(a)), the bond orders will be 1.403. This result matches well with Grabowsky \textit{et al.} with an error of 6.47$\%$ who reported the bond order of sulfur dioxide to be $\sim$ 1.5 from X-ray diffraction data \cite{bondorder}.\\

For the TNI optimized ground state, the bond order comes out to be 0.47 for both the S$_{(1)}$-O$_{(2)}$ and S$_{(1)}$-O$_{(3)}$ double bonds. This decrease in the strength of the bond accounts for the elongation in the bond in Fig. \ref{SO2}(b). 

\begin{table*}[h]
	\centering 		
	\caption{Calculated NBO bond order for atom bonds in the molecule shown in Fig. \ref{nbo2}. }
	\begin{tabular}{c c c c c c} \\ 
		\hline
		\hline
		& Bond & Bond Order & Ref \cite{bondorder}\\ 
		\hline 			
		
		SO$_2$ ground state  & S(1)-O(2) &  0.903 & $\sim $ 1.5 \\
		 & S(1)-O(3) & 1.903 & $\sim $ 1.5 \\
	
		\hline
		SO$_2^-$ ground state & S(1)-O(2) & 0.473 & - \\
		 &  S(1)-O(3) & 0.473 & - \\
		
		%
		\hline
	\end{tabular} \label{bond_order}
\end{table*}

\section{Vibration spectral analysis}
\begin{figure}[h]
	\centering	
	\includegraphics[width=1\linewidth]{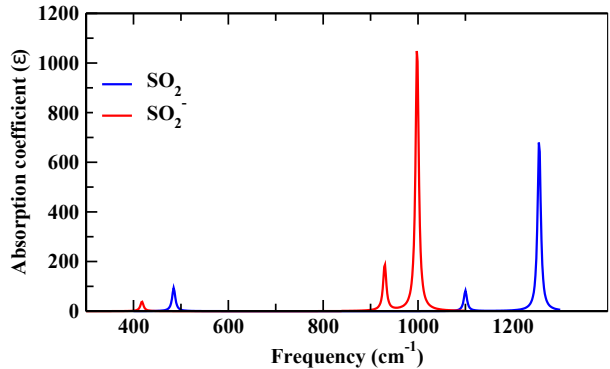}
	\caption{\label{IR} Computed IR spectrum showing different vibration modes for optimized ground state geometry of SO$_2$ molecule (blue solid line) and SO$_2^-$ ion (red solid line). }
\end{figure}

\begin{figure}[h]
	\centering	
	\includegraphics[width=0.8\linewidth]{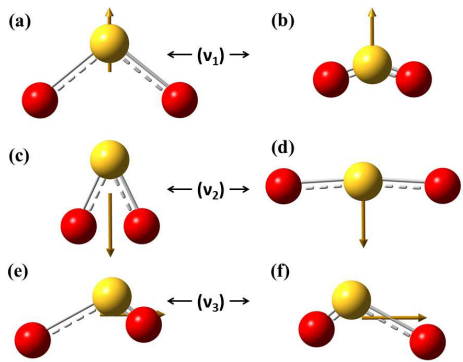}
	\caption{\label{modes} Depiction of computed vibration modes for SO$_2^{-}$ molecule.The yellow sphere represents the S-atom and the red spheres represent the O-atoms. (a) and (b) represent symmetric stretching mode; (c) and (d) represent symmetric bending mode; (e) and (f) represent anti-symmetric stretching mode. The dark yellow arrow denotes the dipole derivative unit vector in all the cases.}
\end{figure}

The main objective of computing the vibration spectra of the neutral SO$_2$ and the TNI SO$_2^{-}$ was to find out the vibration modes connected with the molecular structure. This in turn, helps to identify the modes responsible for the production of negative fragment ions from the TNI and also note the change in the spectra of the negatively charged molecular ion from the neutral parent molecule (Fig. \ref{IR}). When the frequency of radiation matches exactly with the vibration frequency of the molecule, absorption of radiation takes place. Thus, the peaks in the IR spectra denote vibration modes at respective peak positions based on the wave number predicted theoretically by the density functional B3LYP/aug-cc-pVQZ method \cite{boopathi}. The analysis of the vibration spectra was done with the optimized structures of SO$_2$ and SO$_2^{-}$. The three vibrational frequencies for neutral SO$_2$ was computed to be 1167.96 (symmetric stretching, $\nu_1$), 519.22 (symmetric bending, $\nu_2$) and 1357.28 (asymmetric stretching, $\nu_3$) cm$^{-1}$ respectively. While the same for SO$_2^{-}$ was computed to be 979.51 ($\nu_1$), 452.23 ($\nu_2$) and 1071.95 ($\nu_3$) cm$^{-1}$ respectively, as shown in Fig. \ref{modes}. The vibrational frequencies of the three normal modes of vibration for sulfur dioxide were recorded experimentally by Shelton \textit{et al.} using an infrared prism-grating spectrometer \cite{shelton}. The first ($\nu_1$) and third ($\nu_3$) modes both with an $A_1$ symmetry were observed at 1151.38 and 1361.76 cm$^{-1}$, respectively. While the second mode ($\nu_2$) with a $B_1$ symmetry was observed at 517.69 cm$^{-1}$. A. Gordon Briggs also reported the vibrational frequencies of SO$_2$ using double-beam infrared spectrometer fitted with a diffraction grating or NaCl prism to be at 1153 ($\nu_1$), 508 ($\nu_2$) and 1362 cm$^{-1}$ ($\nu_3$) \cite{briggs}. Thus the computed vibrational frequencies of the present work for SO$_2$ matches excellently with literature \cite{shelton,briggs}. But no experimental or theoretical data detailing the vibrational frequencies for SO$_2^{-}$ has been reported till now. The frequencies for the three modes for SO$_2^{-}$ are much less as compared to the frequencies for the three modes for SO$_2$. The change in peak positions for the two before said structures corresponding to different vibration modes are shown in Table \ref{so2_freq}.\\

As can be inferred from Fig. \ref{modes}, the simultaneous presence of the symmetric stretching and bending modes will give rise to S$^-$ fragment formation. As the stretching may produce S$^-$, it has to be followed by the bending mode such that the O-atoms can come close enough to form O$_2$. The same may also produce O$_2^-$. The two pathways can be written as:\\

\begin{equation}
SO_2 + e^-\longrightarrow SO_2^{-*}\longrightarrow 
\begin{cases}
S^-+O_2\\  
O_2^-+S
\end{cases}
\end{equation}

The electron affinity of O$_2^-$ and S$^-$ being EA(O$_2$)=0.451 eV and EA(S)=2.077 eV, cross-section of S$^-$ fragment production is much higher than O$_2^-$ \cite{gupta-baluja}. Whereas, the antisymmetric stretching mode may result in the formation of O$^-$ and SO$^-$ negative ions give by the following equations: \\

\begin{equation}
SO_2 + e^-\longrightarrow SO_2^{-*}\longrightarrow
\begin{cases}
O^-+SO \\
SO^-+O
\end{cases}
\end{equation}

The electron affinity of O$^-$ and SO$^-$ being EA(O)=1.462 eV and EA(SO)=1.125 eV, cross-section of O$^-$ fragment production is slightly higher than SO$^-$ but both are observed \cite{gupta-baluja}. This observation is in good agreement with that reported by Gope \textit{et al.} \cite{gope}.\\

\begin{table*}[h]
	\centering 		
	\caption{Vibrational frequencies in neutral SO$_2$ and SO$_2^{-}$ anion. }
	\begin{tabular}{c c c c c c c c} \\ 
		\hline
		\hline
		& & Calculated &     & Ref\cite{shelton} &     & Ref\cite{briggs}\\ 
			& & (cm$^{-1}$) &     & (cm$^{-1}$) &     & (cm$^{-1}$)\\ 
		\hline 			
		
		DFT Calculation &$\nu_1$ & 1167.96 &     & 1151.38 &      & 1153 \\
		for neutral SO$_2$ & $\nu_2$ & 519.22 &     & 517.69 &      & 508 \\
		& $\nu_3$ & 1357.28 &     & 1361.76 &      & 1362 \\
		\hline
		DFT Calculation &$\nu_1$ & 979.51 &     & - &      & - \\
		for SO$_2^{-}$ anion & $\nu_2$ & 452.23 &     & - &      & - \\
		& $\nu_3$ & 1071.95 &     & - &      & - \\
		
		%
		\hline
	\end{tabular} \label{so2_freq}
\end{table*}

\section{Potential energy curves and charge distribution analysis}

\begin{figure}[h]
	\centering	
	\includegraphics[width=1\linewidth]{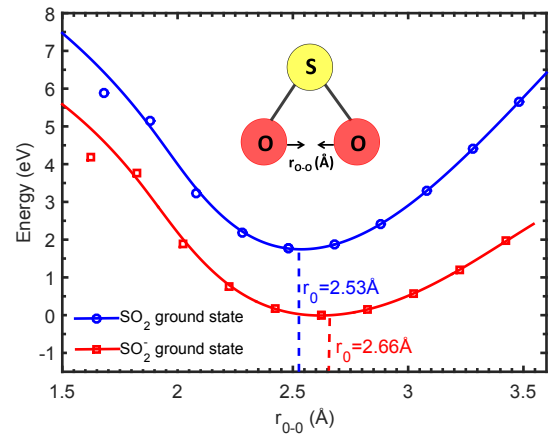}
	\caption{\label{pes1} Variation of potential energy of SO$_2$ (blue solid line) and SO$_2^{-}$ (red solid line) ground states with O-O bond distance. r$_0$ denotes the equilibrium O-O bond distance in both the curves. }
\end{figure}

\begin{figure}[h]
	\centering	
	\includegraphics[width=1\linewidth]{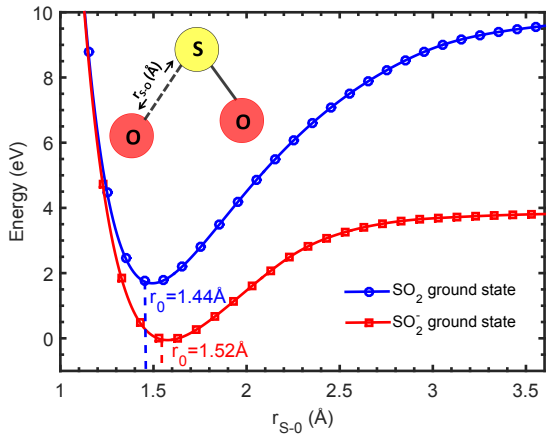}
	\caption{\label{pes2} Variation of potential energy of SO$_2$ (blue solid line) and SO$_2^{-}$ (red solid line) ground states with S-O bond distance. r$_0$ denotes the equilibrium S-O bond distance in both the curves. }
\end{figure}

The potential energy plots with positive adiabatic electron affinity for SO$_2$ and SO$_2^{-}$ ground states have been shown in Fig. \ref{pes1} and Fig. \ref{pes2} for the following dissociation pathways:

\begin{equation}
SO_2 \longrightarrow 
\begin{cases}
O+SO \\
S+O_2  
\end{cases}
\label{eqn4}
\end{equation}

\begin{equation}
SO_2 + e^-\longrightarrow SO_2^{-}\longrightarrow 
\begin{cases}
O^-+SO \\
S^-+O_2  
\end{cases}
\label{eqn5}
\end{equation}
where the B3LYP correlational functional has been used with the aug-cc-pVDZ basis set. To explore the potential energy curves for SO$_2$ and SO$_2^{-}$ optimized ground state geometries for the two dissociation pathways given by Eqns. \ref{eqn4} and \ref{eqn5}, we investigated the transition states and scanned over the bond distances keeping the transition states of SO$_2$ and SO$_2^{-}$ as the initial geometry in two different ways. In the first calculation, the two O-atoms are brought closer from its equilibrium value (denoted by r$_0$ in Fig. \ref{pes1}) and the energy is scanned. This results in the formation of O$_2$ and S fragments from SO$_2$ while, S$^-$ and O$_2$ fragments from SO$_2^{-}$. The resulting potential energy curves for SO$_2$ and SO$_2^{-}$ optimized ground state geometries are shown in Fig. \ref{pes1}. \\

In the second case, keeping one S-O bond and O-S-O angle fixed, the other S-O is stretched from its equilibrium value (denoted by r$_0$ in Fig. \ref{pes2}). This results in the formation of SO and O fragments from SO$_2$ while, O$^-$ and SO fragments from SO$_2^{-}$. The resulting potential energy curves for SO$_2$ and SO$_2^{-}$ optimized ground state geometries are shown in Fig. \ref{pes2}.\\

\begin{figure}[h]
	\centering	
	\includegraphics[width=1\linewidth]{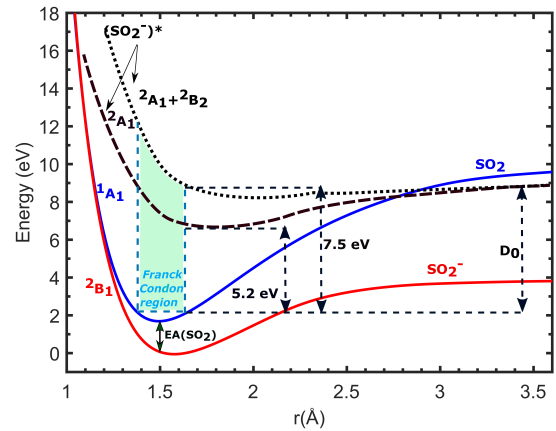}
	\caption{\label{morse} The curve for SO$_2$ (blue solid line) and SO$_2^{-}$ ground state (red solid line) are computed curves for S-O bond distance variation. While, the SO$_2^{-*}$ for the first resonant peak at 5.2 eV (black dashed line) and SO$_2^{-*}$ for the second resonant peak at 7.5 eV (black dotted line) are schematic potential energy curves. D$_0$ signifies the dissociation limit. }
\end{figure}

The DFT calculations predict the potential energy curves of optimized ground state of SO$_2^{-}$ to be at a lower energy than its neutral parent SO$_2$, signifying the positive electron affinity of SO$_2$. But no SO$_2^{-}$ can be observed experimentally to form at around -1.4 eV. This can be understood with the following explanations. As can be noticed from the potential energy curves in Fig. \ref{pes1} and Fig. \ref{pes2}, there is no crossing of the SO$_2$ and SO$_2^{-}$ curves in the Franck-Condon region. Thus this resonant transition of SO$_2$ to SO$_2^{-}$ is not possible. One more reason for not observing any SO$_2^{-}$ formation at around -1.4 eV could be an extremely small lifetime of the SO$_2^{-}$ state, as compared to the time-of-flight which is in the order of $\sim \mu s$ \cite{jana}. The schematic potential energy curves shown in Fig. \ref{morse} depicts the actual SO$_2^{-*}$ states observed experimentally forming the two resonant peaks at 5.2 and 7.5 eV, respectively \cite{jana}. While, the SO$_2$ and SO$_2^{-}$ ground state curves are computed curves for S-O bond distance variation. The energy values for vertical transition within the Franck Condon region are calculated from TDDFT calculations and match excellently well with the experimental values reported by Jana and Nandi \cite{jana}. However, one can use the equation-of-motion coupled cluster (EOMCC) method to compute the potential energy curves for excited states of the TNI, which is beyond the scope of this work.

\begin{figure}[h]
	\centering	
	\includegraphics[width=1\linewidth]{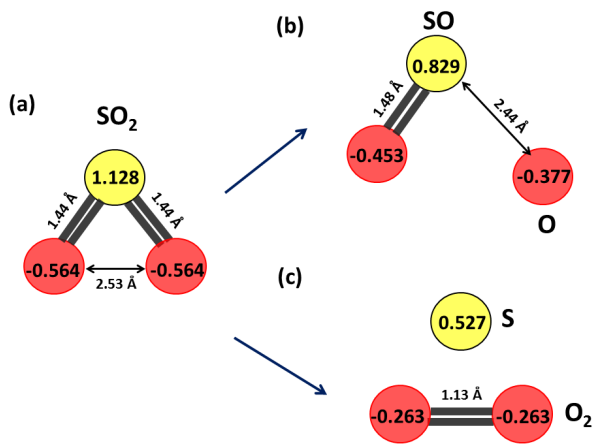}
	\caption{\label{so2_all} Schematic showing Mulliken charge distribution for (a) neutral SO$_2$ molecule; (b) dissociation of the neutral molecule into SO and O fragments; (c) dissociation of the neutral molecule into S and O$_2$ fragments. The yellow sphere denotes S-atom and red spheres denote O-atoms.  }
\end{figure}

The charge distribution of neutral SO$_2$ with equilibrium S-O and O-O bond distances of 1.44$\mathring{A}$ and 2.53$\mathring{A}$ respectively, is shown in Fig. \ref{so2_all}(a), computed using Mulliken charge analysis. Mulliken charges are mathematical constructions that have no relation to physical charges. But it can be used to predict the distribution of charges within the individual atoms forming the molecule before and after dissociation. The total atomic charges add up to give the neutral SO$_2$ molecule although, the S-atom is highly electro-positive and the O-atoms are highly electro-negative. As the S-O bond distance is slowly increased, at some value of the S-O bond coordinate the bond ruptures giving SO and O neutral fragments (Fig. \ref{so2_all}(b)). This value of the bond was computed to be 2.44$\mathring{A}$. From the Mulliken charge analysis it can be observed that both the electro-positivity of the S-atom and the electro-negativity of the O-atoms has decreased in Fig. \ref{so2_all}(b) as compared to \ref{so2_all}(a), predicting the neutral nature of the fragments. Similarly, with the decrease in O-O bond length, S and O$_2$ neutral fragments can be observed to form with 1.13$\mathring{A}$ O-O bond length (Fig. \ref{so2_all}(c)).\\

\begin{figure}[h]
	\centering	
	\includegraphics[width=1\linewidth]{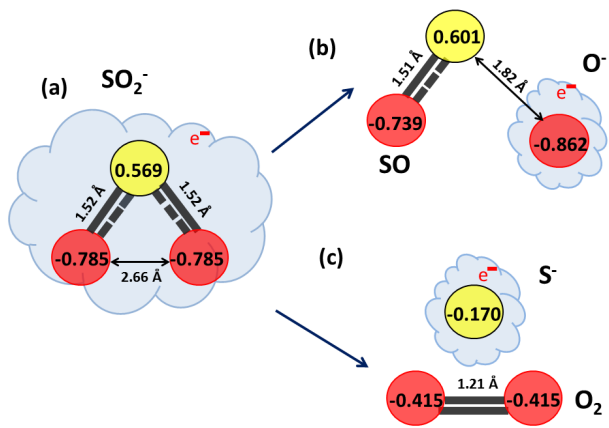}
	\caption{\label{so2-_all} Schematic showing Mulliken charge distribution for (a) TNI SO$_2^{-}$; (b) dissociation of the anion into SO and O$^-$ fragments; (c) dissociation of the anion into S$^-$ and O$_2$ fragments. The yellow sphere denotes S-atom and red spheres denote O-atoms. The blue shaded cloud represents electron cloud.}
\end{figure}

The SO$_2^{-}$ anionic ground state charge distribution is shown in Fig. \ref{so2-_all}(a) with S-O and O-O equilibrium bond distances as 1.52$\mathring{A}$ and 2.66$\mathring{A}$ , respectively. The total charge of this anionic system can be observed to add up -1 unit of electronic charge, signifying the attached low-energy electron. As the S-O bond is gradually stretched, SO and O$^-$ fragment formation can be seen at S-O bond distance of 1.82$\mathring{A}$ (Fig. \ref{so2-_all}(b)). The charge  distribution predicts the SO fragment to be almost neutral, while the O$^-$ fragment carries most of the negative charge. Whereas, gradual decrease in O-O bond distance results in the formation of S$^-$ and O$_2$ fragments, with O-O bond length being 1.21$\mathring{A}$ (Fig. \ref{so2-_all}(c)) which matches excellently with the 1.21$\mathring{A}$ bond length of molecular oxygen.

\section{Symmetry states of the TNI}
In order to get an idea about the symmetry states favorable for DEA to SO$_2$ molecule, the optimized geometries were then used to carry out time-dependent density functional theory (TD-DFT) calculations and get the vertical energy values and symmetries of corresponding excited states for the SO$_2^{-}$ molecule lying within the Franck Condon region. Then the symmetry states lying near the resonant peaks (first resonance observed in the range 4.6 - 5.6 eV and second resonance observed in the range 7.0 - 7.6 eV by different groups \cite{EK1,EK2,jana}) and having symmetries with considerable value of oscillator strength (f) are noted (Fig. \ref{morse}). To check the validity of the calculations, the symmetries are then compared with reported experimental and theoretical data.\\

In TDDFT calculations, oscillator strength (f) denotes the probability of transition between energy levels in an atom or molecule with the help of absorption or emission of electromagnetic radiation. Thus a high oscillator strength for a specific transition would mean that particular transition is more probable than the others. The table showing the symmetry states for the two resonances with considerable f-value are shown in Table \ref{symm} in order of descending f-value.\\

\begin{table*}[h]
	\centering 		
	\caption{Symmetry states of excited TNI with corresponding oscillator strengths in descending order which are expected to contribute to DEA to SO$_2$ from TD-DFT calculations with SO$_2^-$ ground state optimized geometry with the aug-cc-pVQZ basis and B3LYP exchange-correlational functional.}
	\begin{tabular}{c c c c c} \\ 
		\hline
		\hline
		Resonance& Energy & Symmetry & Oscillator & Ref\cite{jana}\\ 
		& range (eV) & of TNI  & strength (f)\\
		
		\hline 	
		First & 4.5 - 5.8 eV & A$_1$  & 0.0515& A$_1$  \\
		
		Peak	&   & B$_1$  & 0.0110 &  \\
		
		\hline
		Second & 7.0 - 7.6 eV & A$_1$  & 0.0516&  	\\ 
		Peak	&   & B$_2$  & 0.0445 & A$_1$ + B$_2$ \\
		&   & B$_2$  & 0.0383 &  \\
		&   & B$_2$  & 0.0319 &  \\
		&   & A$_1$  & 0.0240 &  \\
		&   & B$_2$  & 0.0111 &  \\
		\hline
	\end{tabular} \label{symm}
\end{table*}

As can be noted from the Table \ref{symm}, in the energy range 4.6 - 5.6 eV, the state having the highest oscillator strength is an A$_1$ state with f = 0.0515. Another B$_1$ state can also be observed in the considered energy range but with very low oscillator strength and hence can be neglected. No other symmetry states were found in this energy range with considerable f-value (almost 7.2 times less than 0.05). Thus the computation predicts that the first resonant peak due to DEA to SO$_2$ results from an A$_1$ negative ion resonant state. This matches excellently with previous reports \cite{EK1,EK2,jana}. \\

For the second resonance, negative ion resonant states lying in the energy range 7.0 - 7.6 eV with considerable f - value can be assumed to have a contribution. Six symmetry states can be observed with high f-values and are reported in Table \ref{symm}. Amongst these, four are B$_2$ and two are A$_1$ symmetry states. From this observation, it can be concluded that the second resonance occurs due to an A$_1$ to A$_1$ + B$_2$ transition. This result matches excellently with the recent experimental work on DEA to SO$_2$ by Jana and Nandi \cite{jana}. For all the basis sets used, the oscillator strengths of A$_2$ negative ion states are always 0. This supports the fact that if the initial state of the neutral is an A$_1$, then A$_1$ to A$_2$ transition is always forbidden \cite{EK1}.\\

\section{Conclusion}
The theoretical and computational approach helps to visualize the attachment of a single low-energy ($\leq$ 15 eV) electron to the neutral SO$_2$ molecule thus forming the molecular anion SO$_2^-$. The molecular sites acting as positive centers to the incoming electron could be identified using the MEP of the neutral molecule and the results are in well agreement with the MEP reported by Esrafili and Vakili \cite{Vakili}. The MEP of the TNI also helped to identify the distribution of negative charge in the molecule after the attachment of the electron. Dissociation of the TNI into neutral and charged fragments based on the different vibrational modes have been explained. Computing of the potential energy curves for the dissociation pathways given in Eqn.\ref{eqn4} and Eqn.\ref{eqn5} have been performed for the first time for both the neutral and parent anionic ground states of SO$_2$ molecule. The negative ion resonant states responsible for  the two resonant peaks have also been identified with the help of TD-DFT calculations to the optimized ground state geometries as A$_1$ for the first resonance and A$_1$+B$_2$ for the second resonance, respectively. The results match excellently with that reported experimentally from DEA to SO$_2$ by Jana and Nandi \cite{jana}.

\section{Acknowledgments}
We gratefully acknowledge financial supports from Science and Engineering Research Board (SERB) for supporting this research under the Project EMR/2014/000457 and for computational facility under the Project SB/FTP/CS-164/2013.\\



\section*{References}
\bibliographystyle{iopart-num}
\bibliography{reference}





\end{document}